\documentstyle[epsf,12pt]{article}

\makeatletter

 \@addtoreset{equation}{section}
\makeatother

\begin{document}
\topmargin -35pt
\oddsidemargin 5mm

%%%%%%  user's commands  %%%%%%%%%%%%%%%%%%%%%%%%%%%%%%%%%%%%%%%%%%%
\newcommand {\beq}{\begin{eqnarray}}
\newcommand {\eeq}{\end{eqnarray}}
\newcommand {\non}{\nonumber\\}
\newcommand {\eq}[1]{\label {eq.#1}}
\newcommand {\defeq}{\stackrel{\rm def}{=}}
\newcommand {\gto}{\stackrel{g}{\to}}
\newcommand {\hto}{\stackrel{h}{\to}}
\newcommand {\1}[1]{\frac{1}{#1}}
\newcommand {\2}[1]{\frac{i}{#1}}
\newcommand {\th}{\theta}
\newcommand {\thb}{\bar{\theta}}
\newcommand {\ps}{\psi}
\newcommand {\psb}{\bar{\psi}}
\newcommand {\ph}{\varphi}
\newcommand {\phs}[1]{\varphi^{*#1}}
\newcommand {\sig}{\sigma}
\newcommand {\sigb}{\bar{\sigma}}
\newcommand {\Ph}{\Phi}
\newcommand {\Phd}{\Phi^{\dagger}}
\newcommand {\Sig}{\Sigma}
\newcommand {\Phm}{{\mit\Phi}}
\newcommand {\eps}{\varepsilon}
\newcommand {\del}{\partial}
\newcommand {\dagg}{^{\dagger}}
\newcommand {\pri}{^{\prime}}
\newcommand {\prip}{^{\prime\prime}}
\newcommand {\pripp}{^{\prime\prime\prime}}
\newcommand {\prippp}{^{\prime\prime\prime\prime}}
\newcommand {\pripppp}{^{\prime\prime\prime\prime\prime}}
\newcommand {\delb}{\bar{\partial}}
\newcommand {\zb}{\bar{z}}
\newcommand {\mub}{\bar{\mu}}
\newcommand {\nub}{\bar{\nu}}
\newcommand {\lam}{\lambda}
\newcommand {\lamb}{\bar{\lambda}}
\newcommand {\kap}{\kappa}
\newcommand {\kapb}{\bar{\kappa}}
\newcommand {\xib}{\bar{\xi}}
\newcommand {\ep}{\epsilon}
\newcommand {\epb}{\bar{\epsilon}}
\newcommand {\Ga}{\Gamma}
\newcommand {\rhob}{\bar{\rho}}
\newcommand {\etab}{\bar{\eta}}
\newcommand {\chib}{\bar{\chi}}
\newcommand {\tht}{\tilde{\th}}
\newcommand {\zbasis}[1]{\del/\del z^{#1}}
\newcommand {\zbbasis}[1]{\del/\del \bar{z}^{#1}}
\newcommand {\vecv}{\vec{v}^{\, \prime}}
\newcommand {\vecvd}{\vec{v}^{\, \prime \dagger}}
\newcommand {\vecvs}{\vec{v}^{\, \prime *}}
\newcommand {\alpht}{\tilde{\alpha}}
\newcommand {\xipd}{\xi^{\prime\dagger}}
\newcommand {\pris}{^{\prime *}}
\newcommand {\prid}{^{\prime \dagger}}
\newcommand {\Jto}{\stackrel{J}{\to}}
\newcommand {\vprid}{v^{\prime 2}}
\newcommand {\vpriq}{v^{\prime 4}}
\newcommand {\vt}{\tilde{v}}
\newcommand {\vecvt}{\vec{\tilde{v}}}
\newcommand {\vecpht}{\vec{\tilde{\phi}}}
\newcommand {\pht}{\tilde{\phi}}
\newcommand {\goto}{\stackrel{g_0}{\to}}
\newcommand {\tr}{{\rm tr}\,}
\newcommand {\GC}{G^{\bf C}}
\newcommand {\HC}{H^{\bf C}}
\newcommand{\vs}[1]{\vspace{#1 mm}}
\newcommand{\hs}[1]{\hspace{#1 mm}}

%%%%%%%%%%%%%%%%%%%%%%%%%%%%%%%%%%%%%%%%%%%%%%%%%%%%
\setcounter{page}{0}

%%%%%%%%% title %%%%%%%%%%%%%
\begin{titlepage}

\begin{flushright}
%Ver.7 \\
TIT/HEP-453\\
hep-th/0007184\\
July 2000
\end{flushright}
\bigskip

\begin{center}
{\LARGE\bf
BPS Domain Walls in Models with Flat Directions
}
\vs{10}

\bigskip
{\renewcommand{\thefootnote}{\fnsymbol{footnote}}
{\large\bf Masashi Naganuma\footnote{
     E-mail: naganuma@th.phys.titech.ac.jp .}
 and Muneto Nitta\footnote{
E-mail: nitta@th.phys.titech.ac.jp .}
}}

\setcounter{footnote}{0}
\bigskip

{\small \it
Department of Physics, Tokyo Institute of Technology, 
Oh-okayama, \\ Meguro, Tokyo 152-8551, Japan\\
}

\end{center}
\bigskip

%%%%%%%%% abstract %%%%%%%%
\begin{abstract}
We consider BPS domain walls in four-dimensional ${\cal N}=1$ 
supersymmetric models with continuous global symmetry.
Since the BPS equation is covariant under 
a global transformation,  
the solutions of the BPS walls also have global symmetry.
The moduli space of the supersymmetric vacua in such models 
has non-compact flat directions, and complex BPS walls 
interpolating between two disjoint flat directions can exist. 
We examine this possibility in two models with global $O(2)$ symmetry 
and construct the solutions of such BPS walls. 
\end{abstract}
\vfill
\end{titlepage}

%%%%%%%%%%%%%%%%%%%%%%%%%%
\section{Introduction}
%%%%%%%%%%%%%%%%%%%%%%%%%%%

In recent years, there has been much investigation of domain walls
which appear in many areas of physics. These domain walls interpolate
between degenerate discrete minima of a scalar potential, 
with dependence on one spatial coordinate.
They can occur naturally when a discrete symmetry is 
spontaneously broken.

Domain walls can also appear in supersymmetric 
field theories when the superpotential has more than two critical 
points corresponding to degenerate minima of the scalar potential. 
In particular, it has been found that domain walls in supersymmetric 
theories can saturate the Bogomol'nyi bound. \cite{BPS}  
Such domain walls are called BPS domain walls and preserve half 
of the original supersymmetry. \cite{WittenOlive}
The existence of BPS domain walls 
corresponds to the central extension of ${\cal N}=1$ superalgebra,
and the topological charge of the walls becomes the central charge 
$Z$ of the superalgebra. \cite{Gauntlett}-\cite{DvaliShifman} 
The BPS bound and supercharges are determined 
by this central charge $Z$.

BPS domain walls in supersymmetric theories have been  
extensively studied in models with degenerate isolated vacua. 
\cite{KSS}-\cite{ShifmanVoloshin} 
Moreover, it has been found that such BPS domain walls can 
form a junction when three or more different isolated vacua 
occur in separate regions of space. 
The BPS state of the junction preserves $1/4$ supersymmetry, 
and the BPS bound is determined by two kinds of central charges,
$Z$ and $Y$, appearing in the ${\cal N}=1$ superalgebra. 
\cite{GibbonsTownsend} 
There has been progress in the study of the general properties 
of such BPS junctions,
\cite{Saffin}-\cite{GTT} for example, the negative contribution 
of the charge $Y$ to the junction mass \cite{HKMN} and the 
non-normalizability of zero modes on the BPS junction. \cite{KMHN} 
It has also been argued that BPS junctions can create 
a network \cite{Saffin} 
and that they can play a role in our world 
in higher-dimensional spacetime with a negative cosmological 
constant. \cite{Carroll}  

In this way, 
it has been found that BPS domain walls have many interesting 
properties using models with several {\it isolated} vacua. 
It is essential in these models that isolated vacua have 
different values of the superpotential, 
since their differences are related to 
the energy densities saturating the BPS bound. 
In many supersymmetric theories, 
however, the vacuum manifold consists 
of a {\it continuously degenerated} moduli space. 
Since supersymmetric vacua are 
the extrema of the superpotential ($W'=0$),   
the connected parts of the moduli space have the same values  
of the superpotential, and thus each connected part 
is mapped to a single {\it point} in the superpotential space.  
Hence we can expect the existence of domain walls 
in the models, with the moduli space composed of 
several disjoint parts, rather than isolated points,  
because these disjoint vacua (in the field space)  
are, in general, mapped to {\it different points} 
in the superpotential space. 
In short, the moduli spaces of disjoint supersymmetric vacua 
appear the same as the isolated vacua in the superpotential space.

In this work, we investigate BPS domain walls in ${\cal N}=1$ 
four-dimensional supersymmetric field theories with continuous global 
symmetry. If the models have vacua with spontaneously broken 
global symmetry, there exists a flat direction along the broken symmetry, 
or the moduli space of the vacua. 
Domain walls in such theories can be expected 
to connect pairs of the vacua in the disjoint moduli spaces 
if the Homotopy group $\pi _0$ is nontrivial.\footnote{ 
   Investigation of the Homotopy group gives the {\it necessary} 
   conditions for the existence of domain walls, but there is 
   not always a solution of the equation of motion, in particular
   the BPS equation for the BPS domain walls.
} 
If a BPS domain wall connects such disjoint moduli spaces 
for broken global symmetry,
the configuration itself breaks the symmetry. 
Hence there can be a family of BPS walls 
interpolating between two disjoint moduli spaces;
the BPS bound for walls is given by the difference between the 
superpotential values corresponding to two vacua, and this never
changes under the symmetry transformation.
In fact, we show that 
applying a symmetry transformation to a BPS domain wall solution 
produces another solution of the BPS equation.
Therefore we can expect {\it additional} moduli of BPS walls, 
in addition to the location of the wall's center.   

There is another reason why we study the BPS walls in 
models with continuous global symmetry. 
It is known that when a supersymmetric 
model possesses global symmetry, the superpotential has 
a larger symmetry, 
or the complexification of the original global symmetry, 
owing to the holomorphy of the superpotential.
The vacuum manifold has non-compact flat directions,  
corresponding to the imaginary parts  
of the vacuum expectation values of the fields. 
The Nambu-Goldstone theorem for supersymmetric models 
has been proven.~\cite{KOY} From this, it is known that 
when a global symmetry is spontaneously broken 
in {\it supersymmetric} vacua, there appear NG supermultiplets
as many as the number of the broken generators of the 
complexified group.
Since the complexified group is the symmetry of the superpotential, 
{\it not} that of the whole model nor of the BPS equation,  
it is a highly nontrivial problem to determine whether there can exist 
BPS walls interpolating between two vacua 
along disjoint non-compact flat directions. 
We examine this problem by using 
two supersymmetric models with global $O(2)$ symmetry, 
consisting of two chiral superfields. 
Unlike the global $O(2)$ symmetry, $O(2)^{\bf C}$ transformations
of a BPS wall solution are not solutions of the BPS equation. 
However, we show that there can exist moduli of BPS walls 
corresponding to the shift of vacua along the non-compact flat direction.
This moduli is different from the imaginary part of the parameter 
of the $O(2)^{\bf C}$ transformation. 

In sect.~2, we discuss the general properties of BPS domain walls 
in the model with continuous global symmetry.
In sect.~3, we introduce our two models with $O(2)$ symmetry. 
We examine the existence of complex BPS walls 
interpolating between non-compact flat directions 
in both the models. 
In sect.~4, we reach conclusions for both models  
and discuss the features of BPS domain walls 
in general models with global symmetry. 
We also discuss a possible extension of 
the supersymmetric Nambu-Goldstone theorem. 

%%%%%%%%%%%%%%%%%%%%%%%%%%%%%%%%%%%%%%%%%%%%%%%%%%%%%%%%%%%%%%%%%%%%
\section{BPS walls and continuous symmetry}
%%%%%%%%%%%%%%%%%%%%%%%%%%%%%%%%%%%%%%%%%%%%%%%%%%%%%%%%%%%%%%%%%%%%
We consider supersymmetric field theories with 
only chiral superfields, 
and the K\"{a}hler potential is assumed to be linear: 
$K=\phi\dagg\phi$. 
The supersymmetric vacua are given as the extrema of the
superpotential $W(\phi_k)$, given by 
\beq
\frac{\partial W}{\partial \phi_k}=0, \quad k=1,\cdots K,
\label{SUvac}
\eeq
where the $\phi_k$ are the scalar components of the chiral 
superfields. It is known that,
denoting two solutions of Eq.~(\ref{SUvac}) 
by $\{\phi_k\}_I$ and $\{\phi_k\}_J$, and the corresponding values
of the superpotential by $W_I$ and $W_J$, 
there exists the lower bound of the surface energy density, 
or tension, 
for walls connecting these two vacua expressed by  
\beq
{\cal E} \equiv \frac{\mbox{Energy}}{\mbox{Area}}\geq 2|W_J-W_I|. 
  \label{bpsbound}
\eeq
The BPS wall for which the equality in Eq.~(\ref{bpsbound}) holds 
satisfies the equation
\cite{ChibisovShifman}
\beq
  \partial_z \phi_k = e^{i\alpha}\frac{\partial W^*}{\partial \phi_k^*},  
\label{eqofbps}
\eeq
where $\alpha = \arg (W_J-W_I)$. Here we have considered the wall 
depending on the coordinate $z$.
Equation (\ref{eqofbps}) is called the ``BPS equation``.

If the superpotential $W$ is invariant under the global symmetry $G$,
\beq
 W(\phi) \to W(g \phi) = W(\phi), \hs{10}
 \phi \gto g \phi, \hs{10} g \in G ,
\eeq
where $\phi$ belongs to unitary representation of $G$,  
Eq.~(\ref{SUvac}) is also invariant under $G$:  
\beq
 {\del W(\phi) \over \del \phi_i} \gto 
  g^{-1 T}_{ij} {\del W(\phi) \over \del \phi_j}.
\eeq
Since the superpotential includes only chiral superfields,  
the invariant group $G$ of the superpotential is enlarged 
to its complexification, $\GC$.
It is known that, in addition to the ordinary Nambu-Goldstone bosons 
corresponding to broken $G$ symmetry, 
there appear so-called quasi-Nambu-Goldstone bosons 
corresponding to broken $\GC$ symmetry.\cite{KOY}
With the fermions of their superpartner, 
they constitute massless chiral superfields.  
The vacuum manifold as a $\GC$-orbit is 
parameterized by these massless bosons, and 
the quasi-Nambu-Goldstone bosons just parameterize 
the non-compact flat directions.\footnote{
   It is known that, in the case of the F-term breaking, 
   there must exist at least one quasi-Nambu-Goldstone boson. 
   Then the vacuum manifold inevitably becomes 
   non-compact.\cite{LS}
}
Therefore, in the moduli space of its supersymmetric vacua, 
there exists a non-compact flat direction along the direction 
of the imaginary part of the scalar fields.
 
We can see that the BPS equation (\ref{eqofbps}) is covariant 
under transformation of the global symmetry $G$,
but it is not covariant under the transformation of $\GC$, 
since the BPS equation includes both 
holomorphic and anti-holomorphic fields.
Then, if we can find a solution of Eq.~(\ref{eqofbps}),
configurations obtained through transformation of this 
solution by 
elements of $G$ are also solutions of the BPS equation. 
However, configurations 
obtained through transformations of a solution 
by elements of $\GC$ 
are not generally solutions of the BPS equation. 
Therefore, if the model has more than two disjoint flat directions, 
it is a nontrivial problem to determine whether there exist 
BPS walls interpolating between them. 
We examine this problem 
in two supersymmetric models.

%%%%%%%%%%%%%%%%%%%%%%%%%%%%%%%%%%%%%%%%%%%%%%%%%%%%%%
\section{BPS walls in models with flat directions}
%%%%%%%%%%%%%%%%%%%%%%%%%%%%%%%%%%%%%%%%%%%%%%%%%%%%%%
%%%%%%%%%%%%%%%%%%%%%%%%%%%%%%%%%%%%%%%%%%%%
\subsection{Moduli spaces of our models with flat directions}
%%%%%%%%%%%%%%%%%%%%%%%%%%%%%%%%%%%%%%%%%%%

In this paper, we consider the following two supersymmetric models 
with flat directions.\footnote{
The two models that we consider in this paper are not renormalizable. 
Therefore these models must be interpreted 
as effective theories. 
}
First we consider a model with one flat
direction. Its superpotential is 
\beq
 W(\phi) = \1{4} (\vec{\phi}^{\,2} - a^2)^2, \quad  
 \vec{\phi} = \pmatrix{\phi^1 \cr \phi^2}, 
\eeq
where $\phi^1$ and $\phi^2$ are chiral superfields composing the
doublet of $O(2)$, $\vec{\phi}$, and $a$ is a constant parameter.
By a field redefinition, 
we can take this parameter $a$ to be real and positive 
without loss of generality. 
This model has two disjoint vacua:
\beq
 \mbox{Vac.~I}&&  \vec{\phi} = 0 
      ,\hs{13.3} W = {a^4 \over 4},\non
 \mbox{Vac.~II}&& \vec{\phi}^{\,2} = a^2 
      ,\hs{10} W = 0. 
\label{1vac}
\eeq
Let us note that the $\phi^i$ are the scalar components of chiral 
superfields here. (We denote the chiral superfields and their 
scalar components by the same letter.)
Vac.~I is $O(2)$ symmetric, but Vac.~II spontaneously 
breaks $O(2)$ symmetry. 
The expectation value for Vac.~II can be labeled as 
\beq
 \phi^1 = a \cos \th, \hs{10} 
 \phi^2 = a \sin \th .
\eeq
Now the fields $\phi^1$ and $\phi^2$ can take complex values, 
and we can regard $\th$ as a {\it complex} parameter.
Therefore the vacuum manifold of this model is enlarged 
to an $O(2)^{\bf C}$-orbit:
If we set $\vec{\phi}=\vec{x}+i\vec{y}$, the two disjoint vacua
in the Eq.~(\ref{1vac}) become
\beq
 \mbox{Vac.~I}&&  \vec{x} = \vec{y} = \vec{0}, 
\non
 \mbox{Vac.~II}&& \vec{x}^2 - \vec{y}^2 = a^2, \,\,\, \mbox{and}\,\,\, 
                  \vec{x} \cdot \vec{y} = 0.   
\label{1-vac}
\eeq
\begin{figure}[t]
\begin{center}
%%%%%%%%%%%%%%%%epsfig.sty%%%%%%%%%%%
\leavevmode
\begin{eqnarray*}
\begin{array}{cc}
  \epsfxsize=6cm
  \epsfysize=4.5cm
\epsfbox{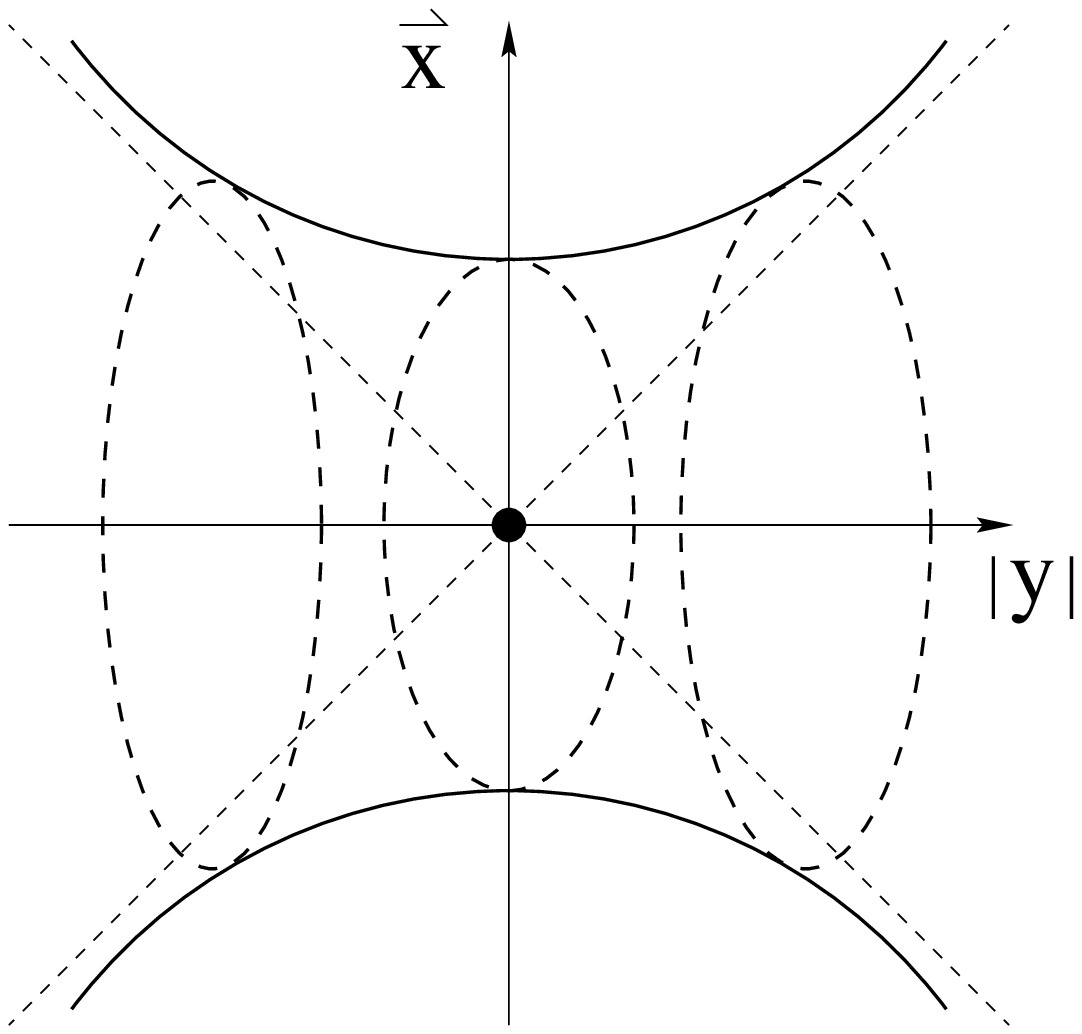} & 
  \epsfxsize=6cm
  \epsfysize=4.5cm
\hspace*{1cm}
\epsfbox{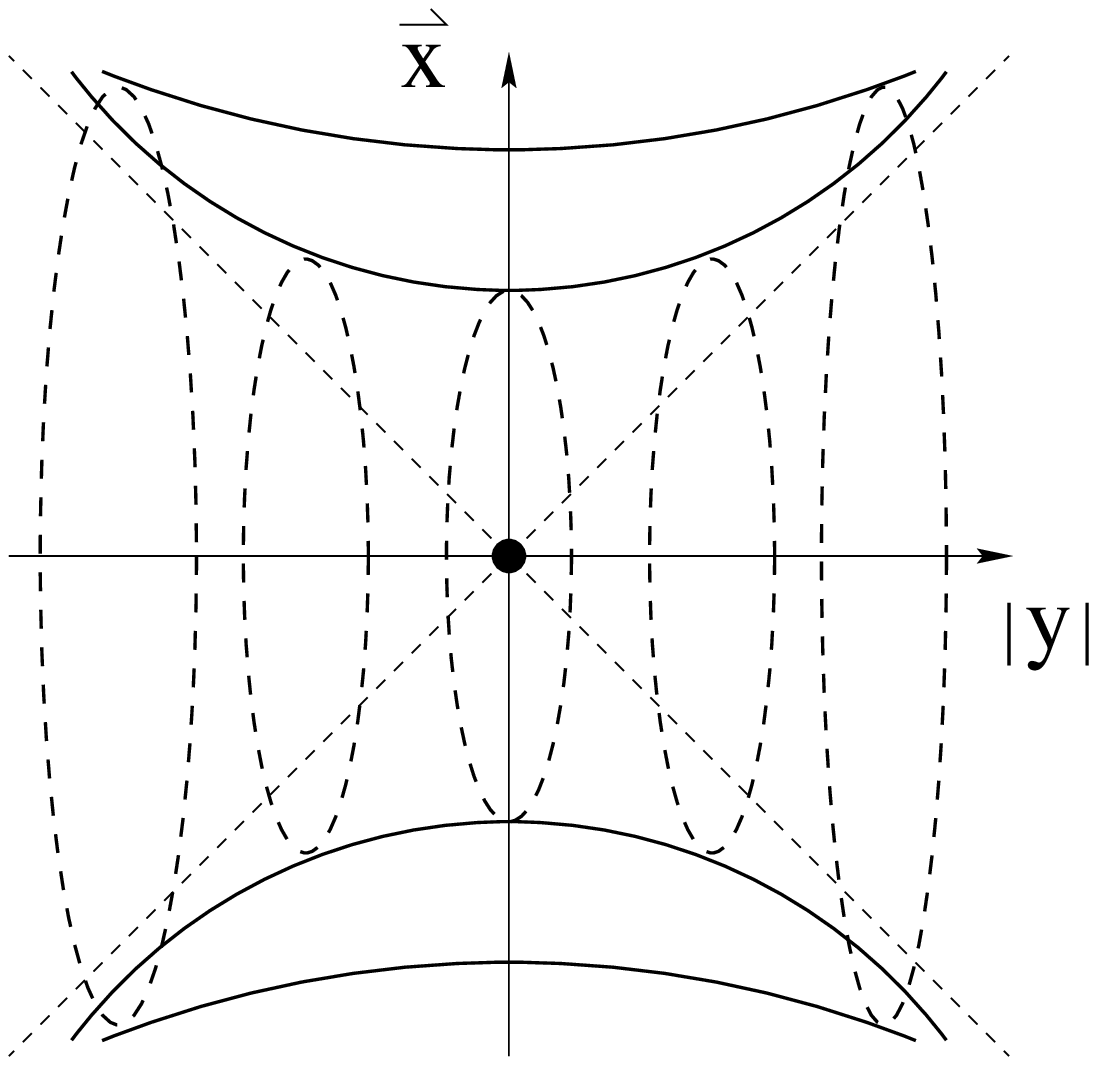} 
\\
\mbox{\footnotesize (a)Moduli space of the first model} & 
\hspace*{1cm}
\mbox{\footnotesize (b)Moduli space of the second model}
\end{array} 
\end{eqnarray*} 
%%%%%%%%%%%%%%%graphics.sty%%%%%%%%%
%%\includegraphics{well.ps}
%%%%%%%%%%%%%%%%%%%%%%%%%%%%%%%%%%%%
\caption{
\footnotesize
(a) The moduli space of the first model is composed of 
two disjoint parts: 
the origin and one hyperboloid. 
The hyperboloid has one compact direction, represented 
by the broken circles,  
and one non-compact direction, represented by the hyperbola.  
(b) The moduli space of the second model is 
composed of three disjoint parts: 
the origin and two hyperboloid with different sizes. 
In both (a) and (b), 
the horizontal axis is $|y|=\sqrt{\vec{y}^2}$, 
the vertical axis is $x^1$, 
and the axis orthogonal to them is $x^2$.
The smallest circle 
corresponds to the real moduli space of vacua, $\vec{y}=0$,
in both figures.
}
\label{MODULI}
\end{center}
\end{figure}
%%%%%%%%%%%%%%%%%%
Hence Vac.~II can be rewritten as a two-dimensional surface 
in the three-dimensional linear space $(x^1,x^2,|y|)$, 
where $|y|=\sqrt{\vec{y}~2}$ (see Fig.~\ref{MODULI}). 
Vac.~II breaks this $O(2)^{\bf C}$ symmetry 
spontaneously.
We consider the BPS wall connecting $O(2)^{\bf C}$ symmetric 
and $O(2)^{\bf C}$ broken vacua, and show that no BPS wall can 
connect the {\it complex vacuum} - the vacuum with a complex value
of the fields shifting along the flat direction in this model 
(see Fig.~\ref{MODULI}).  

Next we consider the model with two flat directions. 
Its superpotential is 
\beq
 W(\phi) = \1{6} \vec{\phi}^{\,2} (\vec{\phi}^{\,2} - a^2)^2, 
\eeq
where $\vec{\phi}$ is an $O(2)$ doublet composed of  
the chiral superfields $\phi^1$ and $\phi^2$,
and the parameter $a$ is assumed to be a positive real constant
for simplicity. 
This model has three disjoint vacua:
\beq
 \mbox{Vac.~I}&&  \vec{\phi} = 0 
      ,\hs{13.5} W = 0 ,\non
 \mbox{Vac.~II}&& \vec{\phi}^{\,2} = {a^2 \over 3}
      ,\hs{9} W = {2 \over 81} a^6, \non
 \mbox{Vac.~III}&& \vec{\phi}^{\,2} = a^2 
      ,\hs{10} W = 0. \label{vac.of_II}
\eeq
Setting $\vec{\phi}=\vec{x}+i\vec{y}$, as in the previous model, 
Vac.~II and Vac.~III can be rewritten as
two hyperboloids with different sizes and Vac.~I as the origin
in the space $(x^1,x^2,|y|)$ (see Fig.~\ref{MODULI}).
 We see that Vac.~I is $O(2)^{\bf C}$ symmetric,  
but Vac.~II and Vac.~III break $O(2)^{\bf C}$ symmetry spontaneously.
We consider the two kinds of BPS walls, connecting Vac.~I and
Vac.~II, and connecting Vac.~II and Vac.~III. 
Then we show that
the BPS walls can connect the complex vacua of Vac.~II and Vac.~III, 
but cannot connect Vac.~I and complex vacua of Vac.~II.

%%%%%%%%%%%%%%%%%%%%%%%%%%%%%%%%%%%%%%%
\subsection{BPS walls in model I}
%%%%%%%%%%%%%%%%%%%%%%%%%%%%%%%%%%%%%%%

Here, we construct BPS saturated walls in the model with one flat 
direction (Model I). 
The BPS equation (\ref{eqofbps}) for this wall is 
\beq
 {\del \phi^i \over \del z} 
 = \phi^{*i} (\vec{\phi}^{*2} - a^2). \label{bpseq-1}
\eeq
First we show that there is no complex solution of this BPS 
equation. 
When we map the field space to the superpotential space, 
two disjoint vacua are mapped to two points.   
It is known that the configuration of the BPS wall can be  
mapped to a line segment 
connecting these two points 
in the superpotential space.~\cite{Fendley} 
Now, the difference between the values of the superpotentials 
for the two vacua, $ \Delta W = a^4 /4$, is real. 
This means that the configuration of the BPS wall in the 
superpotential space is also real. If we set 
$\vec{\phi} = \vec{x} + i \vec{y}$, the imaginary part of the 
superpotential is 
$\Im W = 4 (\vec{x}\cdot\vec{y}) (\vec x^2 - \vec y^2 - a^2)$,  
so we find that BPS solution must satisfy the constraint
$\vec{x}\cdot\vec{y}=0$. 
Using this constraint, the BPS equation of Eq.~(\ref{bpseq-1}) 
can be rewritten as
\begin{eqnarray}
\frac{d}{d z}(\vec{x}+i\vec{y}) 
  = (\vec{x}-i\vec{y})(\vec{x}^2-\vec{y}^2-a^2). 
\label{bpseq-1C}
\end{eqnarray}
From this equation we can derive the following equations:
\begin{eqnarray}
  \frac{d}{d z}\left(\frac{x^2}{x^1} \right)=
  \frac{d}{d z}\left(\frac{y^1}{y^2} \right)=0,\quad 
  \frac{d}{d z}(x^iy^j)=0,\,\, \mbox{for}\,\, i,j=1,2. \label{bpseq-1A}
\end{eqnarray}
The first of these two equations implies that the $O(2)$ rotation 
parameter $\theta$ is constant for the BPS wall. 
Combining these with the constraint $\vec{x}\cdot\vec{y}=0$, 
we can parameterize the BPS wall as
\begin{eqnarray}
  \vec{\phi}(z)  
  =  v(z)\left(
  \begin{array}{c}
    \cos \theta \\
    \sin \theta
  \end{array}
  \right) + i u(z)\left(
  \begin{array}{c}
    \cos (\theta + \pi/2) \\
    \sin (\theta + \pi/2) 
  \end{array}
  \right).  \label{bpseq-1B}
\end{eqnarray}
In Fig.~\ref{MODEL-1} (a), we plot the moduli space of this model 
in the $(u,v)$-plane.
{\footnotesize
\begin{figure}[tbp]
\begin{center}
%%%%%%%%%%%%%%%%epsfig.sty%%%%%%%%%%%
\leavevmode
\begin{eqnarray*}
\begin{array}{cc}
  \epsfysize=5cm
\epsfbox{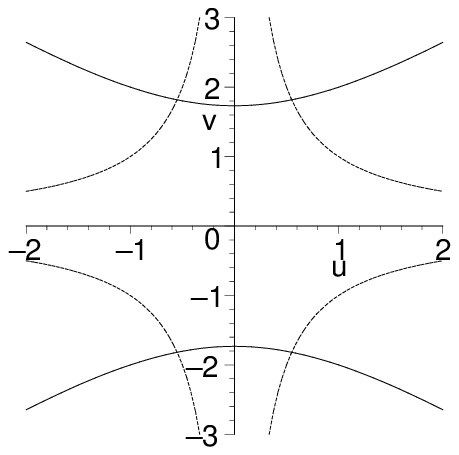} & 
  \epsfysize=5cm
\hspace*{1cm}
\epsfbox{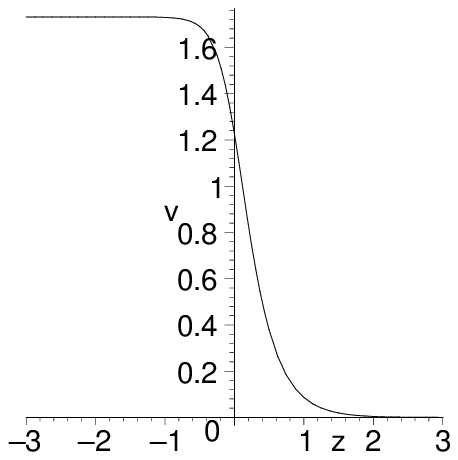} \\
\mbox{(a)Moduli space of model I.} & 
\mbox{(b)Real solution of the BPS domain wall.} 
\end{array} 
\end{eqnarray*} 
%%%%%%%%%%%%%%%graphics.sty%%%%%%%%%
%%\includegraphics{well.ps}
%%%%%%%%%%%%%%%%%%%%%%%%%%%%%%%%%%%%
\caption{
\footnotesize
(a) The moduli space of model I in the $(u,v)$-plane 
is represented by the solid curves, 
and the broken curves correspond to $uv=\pm\sqrt c$ (for $c=1$).
(We set $a^2=3$ in all of the figures in this paper.) 
(b) The real solution, Eq.~(\ref{real-sol-in-I}), 
connecting the origin and 
the nearest points in the hyperbola is 
plotted. 
}
\label{MODEL-1}
\end{center}
\end{figure}
}
%%%%%%%%%%
Substituting Eq.~(\ref{bpseq-1B}) into the second equation 
of Eq.~(\ref{bpseq-1A}),
we can immediately find 
\begin{eqnarray}
  \frac{d(uv)}{dz}=0, \quad uv=\mbox{const}\equiv \sqrt{c},
\end{eqnarray}
where $c$ is a real integral constant. From Fig.~\ref{MODEL-1}(a) 
we find that 
there is no {\it complex} BPS solution connecting Vac.~I and 
vacua along the flat direction of Vac.~II:
In order for a BPS wall to reach Vac.~I, we need to set 
$uv=\sqrt{c}=0$, and 
this is reduced to a real solution [$u(z)=0$] for the boundary 
condition of Vac.~II on the other side.

Hence we consider this solution of Eq.~(\ref{bpseq-1C}). 
This solution can be found as
\beq
 v= \phi_W \equiv a \sqrt{1 \over 1 + \exp
   \left[2a^2(z-z_0)\right]},
 \quad u =0, \label{real-sol-in-I}
\eeq
where $z_0$ is an integral constant, representing 
the position of the center of the domain wall. 
We plot this real solution in Fig.~\ref{MODEL-1}(b). 
Using an $O(2)$ transformation, 
the general real solutions can be written as
\beq
 \phi^1 = \phi_W \cos \th, \hs{10} \phi^2 = \phi_W \sin \th ,
\eeq
where $\th$ is a {\it real} parameter.
The wall separates the two vacua in the broken phase and 
the unbroken phase.
The wall interpolating between the broken and unbroken phase of 
the {\it discrete} symmetry $Z_2$ is discussed in Ref.~\cite{ETBB}. 

%%%%%%%%%%%%%%%%%%%%%%%%%%%%%%%%%%%%%%%
\subsection{BPS wall in the model II}
%%%%%%%%%%%%%%%%%%%%%%%%%%%%%%%%%%%%%%

In this section, we construct BPS walls in the model with two flat 
directions (model II). 
This model has three disjoint vacua 
as in the case of Eq.~(\ref{vac.of_II}). 
The difference between the values of the superpotentials for 
each pair of the three vacua is real, as in the previous model.
There exists no BPS wall connecting Vac.~I and Vac.~III, 
because the two values of the superpotential corresponding to 
these two vacua are 
the same, and the BPS bound (\ref{bpsbound}) becomes zero.  
For this reason we consider two kinds of walls:
walls interpolating between Vac.~II and Vac.~III 
(``outer walls''), 
and walls interpolating between Vac.~I and Vac.~II 
(``inner walls'').
The BPS equations (\ref{eqofbps}) for these walls are 
\beq
 {\del \phi^i \over \del z}  
 = \phi^{*i} \left(\vec{\phi}^{*2} - {a^2 \over 3}\right) 
 (\vec{\phi}^{*2} - a^2),  \label{BPS2}
\eeq
where the boundary conditions are $\vec{\phi}(-\infty)=a^2$ 
[$\vec{\phi}(-\infty)=0$] and $\vec{\phi}(\infty)=a^2/3$ 
for the outer (inner) walls.

The map of the BPS walls into the superpotential space
must be real, as in the previous model:
If we set $\vec{\phi} = \vec x + i \vec y$, 
the imaginary part of the superpotential in this model 
becomes 
\beq
\Im W = \1{3} (\vec{x} \cdot \vec{y})[3(\vec{x}^{\,2}-\vec{y}^{\,2}-a^2)
(\vec{x}^{\,2}-\vec{y}^{\,2}-a^2/3) -4(\vec{x} \cdot \vec{y})^2]. 
\label{ImW}
\eeq
Thus $\vec{x}\cdot \vec{y}=0$ is a sufficient condition.
\footnote{
We can show that this is also a necessary condition 
using the continuity of the solution.}
With this condition, 
Eq.~(\ref{bpseq-1A}) is again valid,
and we can set $\vec{\phi}$ as in Eq.~(\ref{bpseq-1B}).
Hence we can set $\theta = 0 $ in Eq.~(\ref{bpseq-1B}) 
by using the $O(2)$ transformation, without loss of generality, 
yielding $\vec{\phi} = \pmatrix{v \cr iu}$,  
where $v$ and $u$ are real scalar fields.
In Fig.~\ref{MODEL-2A}, 
we illustrate the moduli space of this model 
in the $(u,v)$-plane.
{\footnotesize
\begin{figure}[tbp]
\begin{center}
%\begin{flushleft}
%%%%%%%%%%%%%%%%epsfig.sty%%%%%%%%%%%
\leavevmode
\begin{eqnarray*}
\begin{array}{cc}
  \epsfysize=5cm
\epsfbox{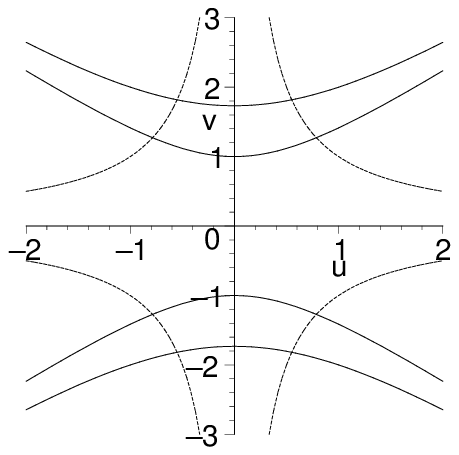} & 
  \epsfysize=5cm
\hspace*{1cm}
\epsfbox{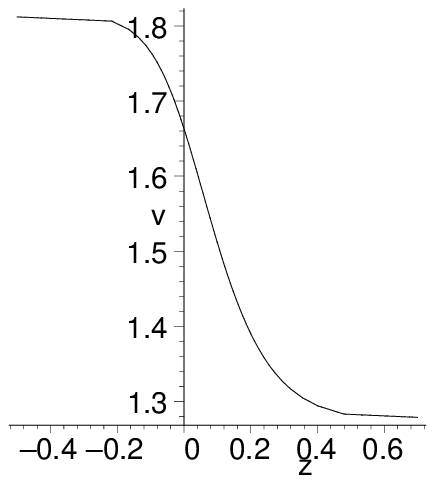} \\
\mbox{(a)Moduli space of model II.} & 
\mbox{(b)Complex solution of the BPS domain wall.} 
\end{array} 
\end{eqnarray*} 
%\centerline{\epsfbox{well.ps}}
%\vspace{5cm}
%%%%%%%%%%%%%%%graphics.sty%%%%%%%%%
%%\includegraphics{well.ps}
%%%%%%%%%%%%%%%%%%%%%%%%%%%%%%%%%%%%
\caption{\footnotesize
(a) The moduli space of model II in the $(u,v)$-plane 
is represented by the solid curves, 
and the broken curves correspond to 
$uv=\pm\sqrt c$ (for $c=1$). 
The value of $v$ for the complex solution, 
Eq.~(\ref{complex_sol2.}), 
connecting between the two hyperbolas along $uv = \sqrt c$ ($c=1$) is 
plotted in (b).    
}
\label{MODEL-2A}
\end{center}
\end{figure}
}
%%%%%%%%%%%%%%%%%%%%%%%%%%%%%%%%%%%%%
Equation~(\ref{BPS2}) becomes 
\begin{eqnarray}
 {d v \over d z} 
 &\!\! =&\!\! v \left(v^2 - u^2 -{a^2 \over 3}\right) 
              (v^2 - u^2 - a^2),\nonumber \\
 {d u \over d z} 
 &\!\! =&\!\! - u \left(v^2 - u^2 -{a^2 \over 3}\right) 
              (v^2 - u^2 - a^2) .
  \label{complex_eq.}
\end{eqnarray}
We can then find 
\beq
 {d(u v) \over dz } = 0.
\eeq
Hence, we can set $uv = \mbox{ const} =\sqrt{c}$. 
We find, from Fig.~\ref{MODEL-2A} (a), that
there can exist a complex BPS wall solution connecting 
Vac.~II and Vac.~III, 
but no complex BPS wall can connect Vac.~I and Vac.~II, 
for the same reason as in model I. 
The first equation in Eq.~(\ref{complex_eq.}) becomes
\beq
 {d v^2 \over d z} 
  = -2 \frac{1}{v^2} \left((v^2)^2 -{a^2 \over 3} v^2 -c \right) 
   ((v^2)^2 - a^2 v^2 -c ).
\eeq
This can be integrated to give
\beq 
e^{-{4a^2\over 3}(z-z_0)}
= \left|{v^2 - \1{2}({a^2 \over 3} + \sqrt{{a^4 \over 9} + 4 c}) 
         \over 
        v^2 - \1{2}({a^2 \over 3} - \sqrt{{a^4 \over 9}+ 4 c}) 
        } \right|^{1 \over \sqrt{{a^4\over 9} + 4c}}
   \left|{v^2 - \1{2} (a^2 - \sqrt{a^4 + 4 c}) \over 
   v^2 - \1{2} (a^2 + \sqrt{a^4 + 4 c}) 
    }\right|^{1 \over \sqrt {a^4 +4c}}, \label{complex_sol1.}
\eeq
where 
$z_0$ is the center of the wall. 
For the complex solution interpolating between Vac.~II and Vac.~III, 
(\ref{complex_sol1.}) can be rewritten as 
\beq
e^{-{4a^2\over 3}(z-z_0)}
= \left[{v^2 - \1{2}({a^2 \over 3} + \sqrt{{a^4 \over 9} + 4 c}) 
         \over 
        v^2 - \1{2}({a^2 \over 3} - \sqrt{{a^4 \over 9}+ 4 c}) 
        } \right]^{1 \over \sqrt{{a^4\over 9} + 4c}}
   \left[{v^2 - \1{2} (a^2 - \sqrt{a^4 + 4 c}) \over 
   \1{2} (a^2 + \sqrt{a^4 + 4 c}) - v^2
    }\right]^{1 \over \sqrt {a^4 +4c}}. \;\;\label{complex_sol2.}
\eeq
Since we cannot obtain an explicit solution $v(z)$ of this equation,
we plot $v(z)$ in the Fig.~\ref{MODEL-2A}(b) 
as an implicit solution of a complex BPS wall. 

We must note that the complex solution of $uv=\sqrt{c}$
is not the $O(2)^{\bf C}$ transformation of the solution of $uv=0$. 
Let us consider a vacuum  
transformed by a $O(2)^{\bf C}$ parameter from a real 
expectation value in Vac.~II. 
The complex BPS wall solution connects this Vac.~II to the 
Vac.~III transformed by a different $O(2)^{\bf C}$ parameter from 
the corresponding real expectation value in Vac.~III.
Therefore the $O(2)^{\bf C}$ transformation of a BPS solution 
does not become a solution of the BPS equation; 
the parameter $c$ which labels the imaginary direction
is not associated with the $O(2)^{\bf C}$ symmetry. 

We can find an explicit solution for real BPS walls. 
For the real solution, the integrated BPS equation can be obtained 
by setting $c=0$ in Eq.~(\ref{complex_sol1.}). We have 
\beq
 X \defeq \exp \left[{4 a^4 \over 3} (z-z_0)\right]
 = {|v^2 - a^2| v^4 \over 
      |v^2 - {a^2 \over 3}|^3} 
 = {|\Phi - a^2| \Phi^2 \over 
      |\Phi - {a^2 \over 3}|^3}, \label{real_sol.}
\eeq
where we have defined $\Phi \defeq v^2 = (\phi^1)^2$. 

We solve this equation in the outer region, 
${a^2 \over 3} \leq (\phi^1)^2 \leq a^2$, 
and the inner region, $0 \leq (\phi^1)^2 \leq {a^2 \over 3}$, 
separately. 

In the case of the outer solutions, 
${a^2 \over 3} \leq (\phi^1)^2 \leq a^2$, 
Eq.~(\ref{real_sol.}) can be rewritten as 
the third order equation
\beq
 (X+1) \Phi^3 - a^2 (X+1)\Phi^2 +{a^4 \over 3}  X \Phi 
 - {a^6 \over 27} X = 0. \label{outer}
\eeq
Thus the third order equation can be solved to yield
\beq
 (\phi^1)^2 = {a^2 \over 3} \left[ 1 
   + \left({1 \over 1+X} 
     + \sqrt{X \over (X+1)^3}\right)^{1\over 3} 
   +\left({1 \over 1+X} 
     - \sqrt{X \over (X+1)^3}\right)^{1\over 3} 
 \right]  \label{outer_sol}
\eeq
for a real solution. 
(The two other solutions are complex and thus inappropriate.)

In the case of the inner solutions, $0 \leq (\phi^1)^2 \leq
{a^2 \over 3}$, Eq.~(\ref{real_sol.}) can be rewritten as 
\beq
 (X-1) \Phi^3 - a^2 (X-1)\Phi^2 +{a^4 \over 3}  X \Phi 
 - {a^6 \over 27} X = 0. 
\eeq
In this case, we must solve this equation 
for each case $X=1$ and $X\neq 1$ separately.
When $X=1$, the solution of this equation is $(\phi^1)^2=a^2/9$, 
and this corresponds to the expectation 
value at the center of the wall ($z=z_0$).
When $X\neq 1$ ($z\neq z_0$), there are three candidates for 
the solution of the outer wall: 
\beq
 && (\phi^1)^2 = {a^2 \over 3} \left[ 1 
   + \left({1 \over 1-X} 
     + \sqrt{X \over (X-1)^3}\right)^{1\over 3}
     \pmatrix{1 \cr e^{{2\pi \over 3}i} \cr e^{-{2\pi \over 3}i}}
 \right.\non 
 &&\left.   \hs{20}
 +\left({1 \over 1-X} 
     - \sqrt{X \over (X-1)^3}\right)^{1\over 3} 
     \pmatrix{1 \cr e^{-{2\pi \over 3}i}\cr e^{{2\pi \over 3}i}}
 \right] .
\eeq
These solutions are not real and positive, so we must choose 
the correct one for the regions $z < z_0$ ($X<1$) and 
$z > z_0$ ($X>1$).
In the region $z > z_0$, 
the first solution is appropriate for the real solution.
In the region of $z < z_0$, 
the third solution is appropriate. 
(In the latter case, the first solution cannot satisfy 
the correct boundary conditions, 
$(\phi^1)^2(-\infty)=0$,
and the second solution tends to infinity in the limit $z \to z_0$.) 
In summary, we obtain the inner wall solution 
by using the third solution
in the left ($z<z_0$) and the first solution in the right ($z>z_0$): 
\beq
(\phi^1)^2 = 
\left\{ 
      \begin{array}{l}
         {a^2 \over 3} \left[ 1 + \left({1 \over 1-X} 
          + \sqrt{X \over (X-1)^3}\right)^{1\over 3}
          -\left(-{1 \over 1-X} 
          + \sqrt{X \over (X-1)^3}\right)^{1\over 3}  \right] 
          \hs{15.2} (z>z_0) \\
         {a^2 \over 3} \left[ 1 + \left({1 \over 1-X} 
          + \sqrt{X \over (X-1)^3}\right)^{1\over 3}e^{-{2\pi \over 3}i}
          +\left({1 \over 1-X} 
          - \sqrt{X \over (X-1)^3}\right)^{1\over 3}e^{{2\pi \over 3}i}
          \right] \,\,(z<z_0) 
      \end{array} \label{inner_sol}
\right.
\eeq         
The profiles of the outer and inner wall solutions are plotted in 
Fig.~\ref{MODEL-2B}.
{\footnotesize
\begin{figure}[ht]
\begin{center}
%\begin{flushleft}
%%%%%%%%%%%%%%%%epsfig.sty%%%%%%%%%%%
\leavevmode
\begin{eqnarray*}
\begin{array}{cc}
  \epsfysize=5cm
\epsfbox{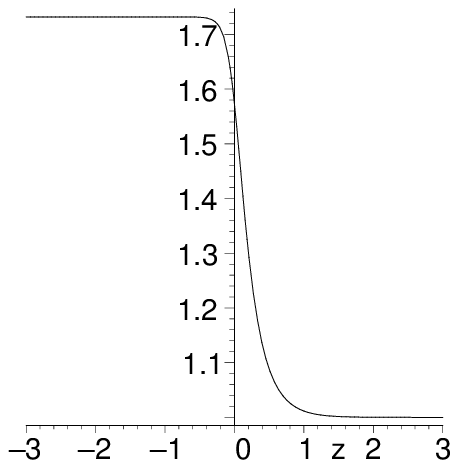} & 
  \epsfysize=5cm
\hspace*{1cm}
\epsfbox{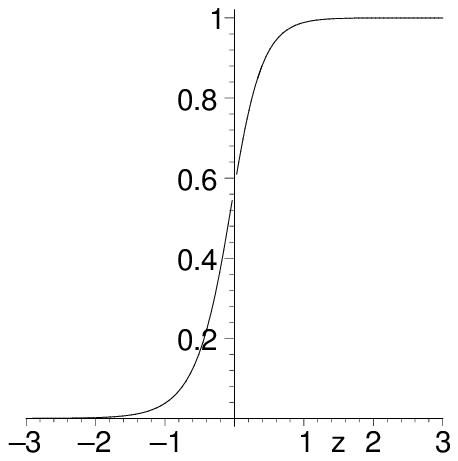} \\
\mbox{(a)Outer solution of the BPS domain wall.} & 
\mbox{(b)Inner solution of the BPS domain wall.} 
\end{array} 
\end{eqnarray*} 
%\centerline{\epsfbox{well.ps}}
%\vspace{5cm}
%%%%%%%%%%%%%%%graphics.sty%%%%%%%%%
%%\includegraphics{well.ps}
%%%%%%%%%%%%%%%%%%%%%%%%%%%%%%%%%%%%
\caption{\footnotesize
The two kinds of real solutions, 
the outer solution, Eq.~(\ref{outer_sol}), 
and the inner solution, Eq.~(\ref{inner_sol}), 
are plotted in (a) and (b), respectively.    
}
\label{MODEL-2B}
\end{center}
%\end{flushleft}
\end{figure}
}
%%%%%%%%%%%%%%%%%%%%%%%%%%%%%%%%%%

%%%%%%%%%%%%%%%%%%%%%%%%%%%%%%%%%%%%%%%%%
\section{Conclusions and discussion}
%%%%%%%%%%%%%%%%%%%%%%%%%%%%%%%%%%%%%%%%

We considered BPS domain walls in models with 
continuously degenerate moduli spaces. 
We discussed only two $O(2)$ symmetric models explicitly,
but many results can be straightforwardly
generalized to other models with a global symmetry $G$. 
When a model has a continuous symmetry, 
$O(2)$ in our models,   
the BPS equation of the wall becomes covariant under this symmetry, 
so the BPS wall also has this symmetry. 
If we can find a BPS solution, 
configurations 
obtained through transformations of this solution 
by elements of $G$ 
are also BPS solutions, 
so they constitute a family of BPS walls. 
Although the boundary conditions change 
under these transformations, 
the tensions of the walls never change.

In supersymmetric field theories, the symmetry $G$ of 
the superpotential is enlarged to its complexification, $\GC$, 
due to the holomorphy of the superpotential.  
Therefore the vacuum manifold includes non-compact flat directions 
corresponding to the directions of imaginary parts of 
the vacuum expectation values. 
As the BPS equation is not covariant under $\GC$,  
it is a highly nontrivial problem 
to determine whether there can exist {\it complex} BPS walls 
interpolating between two disjoint non-compact flat directions. 
To examine this problem,  
we considered two models with flat directions.   
We found that there is no complex BPS wall 
in the first model, 
while there can exist a family of {\it complex} BPS walls 
in the second model. % with two flat directions. 
We have learned from the examination of these two models 
that %, in models with continuous symmetry, 
we must consider {\it complex} configurations for BPS walls 
in models with continuous symmetry.   
This is an important lesson, since 
only real configurations of BPS walls 
have been considered in the literature. 

We have not yet found a criterion to determine 
whether or not a complex BPS wall exists in general models. 
Let us now examine general structures for  
the existence of complex BPS walls  
by counting the number of degrees of freedom 
in these two models. 
Since the BPS equation is a first order differential equation, 
it can be expected that the general solution has 
the same number of integral constants 
as the number of BPS equations,  
{\it unless we enforce the boundary conditions.} 
Since we considered supersymmetric models 
with two chiral superfields, 
there are four BPS equations corresponding to the four real
scalar degrees of freedom. 
However, we have been able to eliminate one degree of freedom, 
since it must be the case of that any BPS solution maps 
to a straight line in the superpotential space. 
We thus can expect that BPS solutions can {\it maximally} 
include three free parameters as the integral constants.
In fact, three parameters, $z_0$, $\theta$ and $c$, 
have appeared as the integral constants in 
the BPS solutions in the second model. 
However, the third parameter $c$ is not 
contained in the BPS wall solution of the first model: 
It was eliminated by the boundary condition. 

We can interpret the parameters $z_0$ and $\theta$ (and $c$),
labeling the solutions of the BPS walls,  
as the ``moduli'' of the BPS wall solutions, 
since the tension of the wall does not change 
when we continuously vary the values of these parameters. 
The configurations obtained under such variations 
are all solutions of a BPS equation, and their tension 
realize the {\it same} BPS bound.
These parameters, however, have slightly different meanings:  
since $z_0$ represents the location of the center of the wall, 
we can vary this parameter 
{\it without} changing the boundary conditions. 
Contrastingly,
we {\it cannot} vary $\theta$ and $c$ without changing 
the boundary condition.

Next we discuss the nature of these parameters 
in terms of symmetry. 
Two of the three parameters represent 
the Nambu-Goldstone modes corresponding 
to the symmetries broken 
by the existence of the wall configuration; 
$z_0$ corresponds to translation along 
the $z$-axis in the spacetime, 
and $\theta$ to the continuous internal symmetry, $O(2)$.  
Therefore the BPS wall solution apparently 
contains these free parameters. 
The parameter $c$ can be considered to represent  
the deformation of the BPS wall along the non-compact flat direction,   
which originates from 
the complexified symmetry of the superpotential. 
This is, however, {\it not} the symmetry of the whole model 
(the K\"{a}hler potential is invariant under $G$ 
but not $\GC$), 
and therefore the additional parameter 
$c$ does not directly correspond to the complex symmetry.   
This is why the BPS wall solutions 
do not always contain $c$. 
Concerning this fact, 
we must comment on the similarity with 
the results in Ref.~\cite{ShifmanVoloshin}. 
As discussed above, 
the parameter $c$ in our model is 
the additional integral constant, 
which depends on the details of the model. 
This quantity is similar to 
the additional integrals of motion  
in Ref.~\cite{ShifmanVoloshin},\footnote{
   Similar additional constants are discussed in the context of 
   non-supersymmetric models in Ref.~\cite{BRS}.
} in the sense that in both models 
the additional constants do not correspond 
directly to the symmetry of the theory. 
However, we must emphasize 
that these quantities have essentially different origins:  
The additional integrals of motion in Ref.~\cite{ShifmanVoloshin} 
represent the spatial distance (in the {\it spacetime}) of 
two separated BPS walls,   
while the quantity $c$ in our model  
controls the shift of the BPS walls along the flat direction 
in the {\it internal} space. 

Let us discuss an interesting problem regarding 
the Nambu-Goldstone theorem 
suggested by our models. 
The moduli space of supersymmetric vacua 
is parameterized by the NG and the quasi-NG bosons 
associated with the spontaneously broken 
$\GC$ symmetry of the superpotential, 
and with their superpartners they constitute 
massless NG chiral multiplets 
as described by the supersymmetric extension of 
the Nambu-Goldstone theorem.~\cite{KOY}  
However, the configuration of the BPS domain wall 
spontaneously breakes half of the supersymmetry 
(and the translational symmetry along the z-axis). 
Therefore, in the entire four-dimensional spacetime, 
${\cal N}=1$ massless NG supermultiplets 
are justified only at infinite distance from the wall.    
The supersymmetric Nambu-Goldstone theorem 
must be deformed around the wall. 
This fact may be a reason why 
the complex parameter of 
the $O(2)^{\bf C}$ transformation 
does not appear as the moduli of BPS wall solutions.  
It would be interesting to examine 
the extension of the Nambu-Goldstone theorem 
to the case of a BPS wall background,  
or the case in which half of the supersymmetry is 
spontaneously broken. 

Before ending this conclusion, 
we point out some interesting features of our models. 
We found that there can exist BPS walls 
connecting $O(2)$ symmetric and $O(2)$ broken vacua. 
(For conventional BPS walls, 
the broken symmetry is usually discrete,  
and vacua separated by the wall 
are {\it both} in the broken phase.) 
Mass spectra are different on opposite sides of 
the walls in our models: 
We can expect massless (quasi-)NG bosons and 
their superpartners only in the broken phase. 
It is a future problem to examine the wave functions of these 
massless modes in order to determine this difference. 

Our second model has three disjoint vacua, and the maps
from two of them to the superpotential space coincide accidentally.
By modifying the model slightly, we can construct a model 
with three disjoint vacua mapped to three distinct points 
in the superpotential space. Hence our examinations can be 
extended to the case of the BPS domain wall junction.

We expect that the new types of BPS domains wall found in 
this paper will play an important role in 
the further understanding of non-perturbative 
aspects of supersymmetric quantum field theories. 

%%%%%%%%%%%%%%%%%%%%%%%%%%%%%%%%%%%%%%%%%%%%%%%%%%%%%%%%%%%%%%%%%%%%%%
\section*{Acknowledgements}
We would like to thank N.~Sakai for valuable discussions. 
We are also grateful to T.~Kugo, H.~Kawai, H.~Kunitomo and 
N.~Sasakura for useful comments.
The work of M.~Nitta is supported in part 
by JSPS Research Fellowships.

%%%%%%%%%%%%%%%%%%%%%%%%%%%%%%%%%%%%%%%%%%%%%%%%%%%%%%%%%%%%%%%%%%%%%%%
%%%%%%%%%%%%%%%%%%%%%%%%%%%%%%%
%%  Reference
%%%%%%%%%%%%%%%%%%%%%%%%%%%%%%

\end{document}